\documentclass[aps,prb,reprint,longbibliography,superscriptaddress]{revtex4-2}
\usepackage{mathrsfs}
\usepackage{amsmath,gensymb}
\usepackage{amsfonts}
\usepackage{amssymb}
\usepackage{amsthm}
\usepackage{graphicx}
\usepackage{natbib}
\usepackage{color}
\usepackage{hyperref}
\usepackage{bm}
\usepackage[caption=false]{subfig}
\usepackage{verbatim}
\usepackage[normalem]{ulem}
\usepackage{soul}

\begin{document}
\title{Gate-Tunable Superconducting Spin Valve in a van der Waals Ferromagnet/Superconductor/Ferromagnet Trilayer}

\author{A. S. Ianovskaia}
\affiliation{Moscow Institute of Physics and Technology, Dolgoprudny, 141700 Moscow region, Russia}

\author{G. A. Bobkov}
\affiliation{Moscow Institute of Physics and Technology, Dolgoprudny, 141700 Moscow region, Russia}

\author{A. M. Bobkov}
\affiliation{Moscow Institute of Physics and Technology, Dolgoprudny, 141700 Moscow region, Russia}

\author{I.V. Bobkova}
\affiliation{Moscow Institute of Physics and Technology, Dolgoprudny, 141700 Moscow region, Russia}
\affiliation{National Research University Higher School of Economics, 101000 Moscow, Russia}

\begin{abstract}

We theoretically demonstrate a gate-tunable superconducting spin valve effect (SVE) in a van der Waals (vdW) heterostructure composed of a monolayer superconductor (S) sandwiched between two ferromagnetic (F) monolayers (F/S/F). By electrostatically gating the ferromagnetic layers to modulate their chemical potentials, the system can be continuously tuned between the \textit{standard}, \textit{inverse} and \textit{triplet} (non-monotonic) SVE regimes within the same device. This tunability originates from the gate-controlled hybridization between the superconducting and ferromagnetic electronic spectra, which determines the effective exchange field induced in the S-layer. Furthermore, we reveal that gating enables exotic, non-BCS temperature dependencies of the superconducting order parameter, including reentrant superconductivity, bistable states, first-order phase transitions, and the emergence of superconductivity at finite temperatures. Our results establish vdW F/S/F trilayers as a versatile and highly controllable platform for superconducting spintronics, where external gate voltages can selectively activate different spin-valve functionalities and unconventional superconducting states.

\end{abstract}

\maketitle

\section{Introduction}

Superconductor/ferromagnet (S/F) heterostructures are a subject of intense research in superconducting spintronics, primarily driven by the unique proximity effects that emerge at their nanoscale interfaces \cite{Buzdin2005,Bergeret2005,Eschrig2015,Linder2015}. One of key device concepts in this field is the superconducting spin valve, which can be implemented in both F/S/F and F/F/S geometries. Its operating principle hinges on the spin-valve effect (SVE), where the superconducting critical temperature $T_c$  is controlled by the relative orientation of the magnetizations in the ferromagnetic layers.

In its simplest form, the SVE can be qualitatively explained as follows. In a thin-film S/F bilayer, the magnetic proximity effect induces an effective exchange field in the superconductor, which suppresses superconductivity \cite{Bergeret2001,Bergeret2018review}. Here, ``thin-film'' refers to a superconducting layer thickness \( d_S \) smaller than the superconducting coherence length \( \xi_{\rm cl} \), although \( d_S \) and \( \xi_{\rm cl} \) may extend to hundreds of monolayers. In an F/S/F trilayer, both ferromagnets induce exchange fields in the superconducting interlayer. These fields add up in the parallel (P) magnetization configuration, but partially compensate each other in the antiparallel (AP) configuration. Consequently, superconductivity is more strongly suppressed in the parallel case, resulting in \( T_c^P < T_c^{AP} \)—the so-called \emph{standard} spin-valve effect \cite{DeGennes1966}.

Most theoretical studies predict the standard SVE in thin-film F/S/F and F/F/S heterostructures, both in the diffusive \cite{Tagirov1999,Buzdin1999,Baladie2001,Baladie2003,Fominov2003,Cadden-Zimansky2008,Mironov2014,Oh1997} and ballistic \cite{Baladie2001,Bozovic2005,Halterman2005,Bobkov2025_valve} transport regimes. Nevertheless, it has been theoretically predicted that the proximity effect in a ballistic thin-film F/S/F structure can also give rise to the inverse SVE \cite{Mironov2014}. In such systems, the critical temperature difference \( T_c^P - T_c^{AP} \) oscillates as a function of the thickness of one ferromagnetic layer, leading to alternating regions where the standard or inverse effect is dominant. These oscillations stem from the characteristic oscillatory decay of the Cooper pair wave function within a ferromagnetic metal \cite{Buzdin2005}.  Significantly, this oscillatory behavior vanishes in the diffusive limit, where Usadel equations consistently yield only the standard SVE \cite{Mironov2014}. Complementary theoretical work by Fominov \textit{et al.} \cite{Fominov2010} demonstrated that both standard and inverse SVE manifestations in F1/F2/S systems emerge from constructive or destructive interference of Cooper pairs at the F1/F2 and F2/S interfaces.

On the experimental side, the standard SVE has been widely observed \cite{Deutscher1969,Nowak2008,Cadden-Zimansky2008,Gu2002,Potenza2005,Moraru2006_M,Moraru2006_O,Zhu2010,Kamashev2024,Gu2015,Li2013,Westerholt2005,Kamashev2025,Bhakat2025,Matsuki2025,DiBernardo2019,Kikuta2024,Komori2025}, including a recent report on the absolute spin valve effect \cite{Matsuki2025}. Nonetheless, several studies have reported clear signatures of the \emph{inverse} SVE, where \( T_c^P > T_c^{AP} \) \cite{Rusanov2006,Steiner2006,Singh2007,Kim2007,Leksin2009,Zhu2009,Banerjee2014,Сolangelo2025,StoddartStones2022,DiBernardo2019,Komori2025}. It is important to note that the physical mechanisms proposed for these experimental observations are not always rooted in the proximity effect. For instance, the inverse SVE was attributed to quasiparticle accumulation in the superconducting layer for the antiparallel configuration in Refs.~\cite{Rusanov2006,Singh2007}, while Refs.~\cite{Steiner2006,Kim2007,Zhu2009} ascribed it to the influence of magnetic stray fields arising from domain walls in the ferromagnets.

Beyond the standard and inverse SVE, the dependence of \( T_c \) on the misorientation angle \( \theta \) can be nonmonotonic, with a minimum near \( \theta = \pi/2 \) \cite{Fominov2010,Mironov2014,Wu2012,Leksin2012,Jara2014,Singh2015,Flokstra2015,Karminskaya2011,Wu2012,Kamashev2024_FFS,Lenk2016_Tr,Zdravkov2013,Zdravkov2013_Memory}. This behavior arises from the generation of equal-spin triplet pairs, which correspond to long-range triplet correlations inside the ferromagnets. These pairs open an additional channel for superconductivity suppression in non-collinear magnetization configurations. Since the triplet correlations are proportional to the cross product of the two magnetizations, they are most efficiently generated near \( \theta = \pi/2 \), leading to a characteristic minimum in the \( T_c(\theta) \) dependence.

In addition to the well-studied thin-film structures, systems comprising a few monolayers—fabricated from van der Waals (vdW) materials \cite{Geim2013,Novoselov2016}—have emerged as promising platforms for investigating proximity effects \cite{Zollner2025}. The weak interlayer coupling in vdW materials allows for layer-by-layer assembly of heterostructures with tailored parameters. Numerous vdW materials are now available, including magnets (e.g., \(\text{Fe}_3\text{GeTe}_2\) \cite{Gibertini2019,Zhang2001,Zhuang2016,Liu2017,Wang2017,Chen2013,Deng2018,Yi2016,Fei2018}, \(\text{VSe}_2\) \cite{Bonilla2018,Ma2012}, \(\text{CrTe}_2\) \cite{Zhang2021}, \(\text{V}_5\text{S}_8\) \cite{Zhang2020}, \(\text{VS}_2\) \cite{Ma2012}) and superconductors (e.g., \(\text{NbSe}_2\) \cite{Xi2015,Soto2006}, \(\text{MoS}_2\) \cite{Saito2016}). Several theoretical \cite{Wickramaratne2021,Aikebaier2022,Bobkov2024_vdW,Ianovskaia2024,Bobkov2024_spin,Bobkov2025} and experimental \cite{Jo2023,Jiang2020,Kezilebieke2020,Ai2021,Idzuchi2021} studies have explored S/F vdW heterostructures. In particular, Refs.~\cite{Bobkov2024_vdW} and \cite{Ianovskaia2024} have theoretically investigated the proximity effect in S/F structures with single or few superconducting monolayers, revealing counterintuitive non-monotonic dependence of superconductivity on the internal exchange field and gate voltage applied to the ferromagnet.

Several theoretical works have been dedicated to the SVE in heterostructures consisting of a few monolayers. Standard, inverse, and triplet SVE have been reported depending on the material and geometric parameters of the trilayer \cite{Buzdin2003,Tollis2005,Montiel2009,Devizorova2017,Devizorova2019}. In this work, we demonstrate that vdW heterostructures provide a versatile platform for implementing all types of SVE—standard, inverse, and triplet—in the \emph{same} structure. The type of SVE can be controlled by a gate potential applied to the \(\rm F_1\) and \(\rm F_2\) layers of the \(\rm F_1/S/F_2\) heterostructure, which adjusts the chemical potentials \(\mu_{F_{1(2)}}\) of the respective ferromagnetic layers.  

By considering a minimal model of the vdW trilayer, we study the full phase diagram of the system in the \((\mu_{F_1},\mu_{F_2})\)-plane, revealing regions of standard, inverse, and triplet SVE. The physical mechanism underlying the controllability of the SVE is analyzed. Moreover, we demonstrate that in the considered \(\rm F_1/S/F_2\) vdW heterostructure, the temperature dependence of the superconducting order parameter (OP) is also widely tunable by gating, even in the most easily realizable experimental case of parallel magnetizations. Depending on the gate voltages applied to the \(\rm F_1\) and \(\rm F_2\) layers, superconductivity can exhibit exotic behavior that is strongly different from standard BCS-like suppression. In particular, we obtain reentrant superconductivity, regions of bistable behavior, appearance of superconductivity at finite temperatures, and first-order superconducting transitions.

The paper is organized as follows. In Sec.~\ref{sec:model}, we introduce the theoretical model for the heterostructure. Sec.~\ref{sec:method} outlines the Green's function formalism employed in our calculations. We present our findings on the gate-tunable standard and inverse SVE and discuss the underlying physical mechanism in Sec.~\ref{sec:St_Inv}. The triplet spin-valve effect is analyzed in Sec.~\ref{sec:Triplet}. Sec.~\ref{sec:OP_T} explores the potential for realizing nontrivial, non-BCS temperature dependencies of the superconducting order parameter through gating. Finally, Sec.~\ref{sec:conclusions} presents the conclusions of our work. Technical details concerning the derivation of the Gor'kov equation for the Green's function are provided in Appendix~\ref{sec:App_Gorkov}, while the additional numerical results for the temperature dependence of the order parameter are presented in Appendix~\ref{app:Delta_T}.

\section{Model}

\label{sec:model}

\begin{figure}[tb]
	\begin{center}
		\includegraphics[width=75mm]{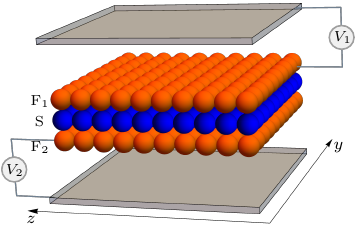}
\caption{\textbf{$\mathbf{F_1/S/F_2}$ van der Waals spin valve.} The heterostructure comprises a monolayer superconductor (S) encapsulated between two ferromagnetic monolayers ($\rm F_1$ and $\rm F_2$). Schematic representation includes top and bottom gate voltages $V_{1,2}$ applied to the respective $\rm F_{1,2}$ layers.}
 \label{fig:sketch}
	\end{center}
 \end{figure}

The system under investigation is schematically depicted in Fig.~\ref{fig:sketch}. It comprises a monolayer superconducting sheet sandwiched between two monolayer ferromagnetic films, forming a prototypical $\rm{F_1/S/F_2}$ superconducting spin valve. The system is modeled by a tight-binding Hamiltonian on a square lattice:
\begin{align}
H = H_{F_1} + H_{F_2} + H_{S} + H_{FS} ,
\label{ham}
\end{align}
where the Hamiltonian for the ferromagnetic layer $\mathrm{F}_n$ ($n=1,2$) is given by
\begin{align}
H_{F_n} = - t_F &\sum_{\langle \bm i \bm j \rangle, \sigma} c^{F_n\dagger}_{\bm i \sigma} c^{F_n}_{\bm j \sigma} - \mu_{F_n} \sum_{\bm i, \sigma} c^{F_n\dagger}_{\bm i \sigma} c^{F_n}_{\bm i \sigma} + \nonumber \\
& \sum_{\bm i, \alpha, \beta} c^{F_n\dagger}_{\bm i \alpha} (\bm h_n \cdot \bm \sigma)_{\alpha \beta} c^{F_n}_{\bm i \beta} ,
\label{ham_F}
\end{align}
the superconducting layer Hamiltonian reads
\begin{align}
H_{S} = - t_S &\sum_{\langle \bm i \bm j \rangle, \sigma} c^{S\dagger}_{\bm i \sigma} c^{S}_{\bm j \sigma} -  \mu_S \sum_{\bm i, \sigma} c^{S\dagger}_{\bm i \sigma} c^{S}_{\bm i \sigma} + \nonumber \\
& \sum_{\bm i} (\Delta c^{S\dagger}_{\bm i \uparrow} c^{S\dagger}_{\bm i \downarrow} + \text{h.c.}),  
\label{ham_S}
\end{align}
and the interlayer coupling is described by
\begin{align}
H_{FS} = - t_{FS} \sum_{\bm i, \sigma, n} (c^{F_n\dagger}_{\bm i \sigma} c^{S}_{\bm i \sigma} + \text{h.c.}).
\label{ham_int}
\end{align}
Here, $c^{S}_{\bm i,\sigma}$ ($c^{F_n}_{\bm i,\sigma}$) denotes the electron annihilation operator in the S ($\mathrm{F}_n$) layer at site $\bm i$ with spin $\sigma = \uparrow, \downarrow$. The parameters $\mu_S$ and $\mu_{F_n}$ represent the onsite energies of the S and $\mathrm{F}_n$ layers, respectively, measured from the bottom of their corresponding conduction bands. For an isolated layer, the onsite energy coincides with its chemical potential. We consider only nearest-neighbor hopping, with $t_S$ ($t_F$) being the intralayer hopping integral within the S ($\mathrm{F}_n$) material. The notation $\langle \bm i \bm j \rangle$ indicates summation over nearest-neighbor sites, and $t_{FS}$ quantifies the interlayer hopping between S and $\mathrm{F}_n$ layers. The exchange field $\bm h_n$ in layer $\mathrm{F}_n$ is taken as spatially uniform within the plane. The ferromagnetic layers are assumed identical with $h_1=h_2=h$, though our results can be readily extended to accommodate different magnetization magnitudes $h_1 \neq h_2$, as discussed in Sec.~\ref{sec:St_Inv}.   The superconducting OP $\Delta$, assumed to be of spin-singlet $s$-wave type and nonzero only in the S layer, is determined self-consistently via $\Delta = \lambda \langle c^{S}_{\bm i \downarrow} c^{S}_{\bm i \uparrow} \rangle$, where $\lambda$ is the pairing constant.

The magnetization vectors $\bm h_1$ and $\bm h_2$ are confined to the $yz$ plane, with an angle $\theta$ between them. Their orientations in spin space are defined as:
\begin{align}
\bm h_{1,2} = h (0, \pm \sin(\theta/2), \cos(\theta/2))^T.
\label{eq:h_n}
\end{align}
We neglect orbital effects arising from stray fields of the ferromagnetic layers. Such fields are predominantly confined to the edges; away from the edges, they are suppressed by a factor of $d/L$ for any magnetization orientation relative to the S/\(\mathrm{F}_n\) interfaces, where $d$ is the film thickness (on the atomic scale) and $L$ denotes the characteristic in-plane dimension of the structure.

We set $t_F/t_S = 1.25$ and $t_{FS} \ll t_{F,S}$. This parameter choice captures the essential qualitative characteristics of the electronic spectra in vdW materials near the Fermi level. Density functional theory (DFT) calculations \cite{Wang2021}, for instance, indicate that interlayer hopping in $\mathrm{NbSe_2}$ is roughly an order of magnitude smaller than intralayer hopping. Moreover, hopping between dissimilar materials is expected to be further reduced \cite{Bobkov2024_vdW} because of lattice mismatch and interface imperfections. 

The primary objective of this work is to explore the gate tunability of the superconducting state in the spin valve. In mono- or few-layer vdW materials, gating provides an effective means to modulate the chemical potential \cite{Xi2016,Deng2018,Matsuoka2024}. Accordingly, within our model, the ferromagnetic onsite energies $\mu_{F_{1,2}}$ are treated as external parameters, with their variations proportional to the applied top ($V_1$) and bottom ($V_2$) gate voltages, respectively.

It is important to note that in the framework of our model the spin and coordinate spaces are independent. Consequently, the plane spanned by the magnetization vectors \(\bm h_1\) and \(\bm h_2\) can be chosen arbitrarily with respect to the S/\(\mathrm{F}_n\) interfaces. However, this model is relatively simple and omits certain realistic features of van der Waals materials, such as Ising or Rashba spin-orbit coupling (SOC). The influence of Ising SOC on the proximity effect in van der Waals superconductor/ferromagnet bilayers (e.g., NbSe$_2$/VSe$_2$) was investigated in detail in Ref. \cite{Bobkov2024_vdW}, including gate-voltage control of the induced exchange field and the resulting suppression of superconductivity. It was shown that while the magnitude of the order parameter suppression depends on the orientation of the magnetization $\mathbf{M}$ relative to the interface—superconductivity is less suppressed for in-plane $\mathbf{M}$ because of Ising protection against Zeeman depairing—the key qualitative feature, namely the electrically tunable suppression and recovery of the order parameter, remains robust. This robustness originates from the fact that the underlying mechanism is governed primarily by the hybridization of the normal-state electronic spectra of the superconductor and ferromagnet layers. The same holds for the triplet correlations induced in the superconductor by the ferromagnet: they are significantly larger for out-of-plane $\mathbf{M}$ than for in-plane $\mathbf{M}$; however, the electrical controllability persists.

Thus, the main distinction between our minimal model (without SOC) and realistic van der Waals materials with Ising SOC is that in the simplified model, only the relative orientation of the two magnetizations matters, while their orientation with respect to the superconductor/ferromagnet interfaces is irrelevant. In the presence of Ising SOC, the orientation of each magnetization relative to the interface becomes an additional crucial parameter. We discuss its influence on the specific type of spin-valve effect in the corresponding sections below.

Ising SOC, in combination with the Zeeman field, also gives rise to nonunitary superconducting correlations, which are essential for dissipationless spin transport \cite{Bobkov2024_spin}. Rashba SOC can be equally important for phenomena such as finite-momentum pairing (the Fulde–Ferrell–Larkin–Ovchinnikov state) \cite{Kaur2005,Akbari2022,Zhang2022,Zhao2023,Wan2023,Ding2023}.

\section{Green's functions technique for F/S/F vdW heterostructure}

\label{sec:method}

We introduce the Nambu spinor 
\begin{widetext}
\begin{eqnarray}
\check \psi_{\bm i} = (c_{{\bm i}\uparrow}^{F_1}, c_{\bm i\downarrow}^{F_1}, c_{\bm i\downarrow}^{F_1\dagger}, -c_{\bm i\uparrow}^{F_1\dagger}, c_{{\bm i}\uparrow}^{S}, c_{\bm i\downarrow}^{S}, c_{\bm i\downarrow}^{S\dagger}, -c_{\bm i\uparrow}^{S\dagger},c_{{\bm i}\uparrow}^{F_2}, c_{\bm i\downarrow}^{F_2}, c_{\bm i\downarrow}^{F_2\dagger}, -c_{\bm i\uparrow}^{F_2\dagger})^T 
\label{eq:spinor}
\end{eqnarray}
\end{widetext} 
and define the Green's function—a $12\times12$ matrix in the direct product of spin, particle-hole, and layer spaces—as follows: 
\begin{eqnarray}
\check G_{\bm i \bm j}(\tau_1, \tau_2) = - \tau_z \langle T_\tau \check \psi_{\bm i}(\tau_1) \check \psi_{\bm j}^\dagger(\tau_2) \rangle,
\label{eq:Green_Gorkov}
\end{eqnarray}
where $\langle T_\tau ... \rangle$ denotes imaginary time-ordered thermal averaging. Throughout this work, we employ Pauli matrices $\sigma_k$ and $\tau_k$ ($k=0,x,y,z$) in spin and particle-hole spaces, respectively. 

Owing to the translational invariance along the S/F interface, we introduce the Fourier-transformed Green's function:
\begin{eqnarray}
\check G(\bm p, \tau) =  \int d^2 r e^{-i \bm p(\bm i - \bm j)}\check G_{\bm i \bm j},
\label{eq:mixed}
\end{eqnarray}
with $\tau = \tau_1 - \tau_2$ and integration performed over $\bm i - \bm j$.

Expanding $\check G(\bm p, \tau)$ in fermionic Matsubara frequencies $\omega_m = \pi T(2m+1)$ as $\check G(\bm p, \tau) = T \sum_{\omega_m} e^{-i \omega_m \tau} \check G(\bm p, \omega_m)$, where $T$ is the temperature, we derive the Gor'kov equation for the Green's function (see Appendix \ref{sec:App_Gorkov} for details):
\begin{widetext}
\begin{align}
    \check G^{-1} \check {G}(\bm p, \omega_m)=1,
\label{eq:Gor'kov_equation}
\end{align}
\begin{equation}
\check{G}^{-1} = \left(\begin{matrix} i \omega_m \tau_z -\xi_{F_1} - \bm h_{1} \bm \sigma \tau_z & t_{FS} &0  \\ t_{FS} & i \omega_m \tau_z -\xi_S + i \Delta \tau_y &  t_{FS}  \\  0&t_{FS}&i \omega_m \tau_z -\xi_{F_2} - \bm h_{2} \bm \sigma \tau_z \\\end{matrix}\right),
\label{eq:matrix_hamiltonian}
\end{equation}
\end{widetext}
where 
\begin{equation}
\xi_S = -2 t_S(\cos p_y a+\cos p_z a) - \mu_S
\label{dispersion_S}
\end{equation}
and 
\begin{equation}
\xi_{F_{1(2)}} = -2 t_{F}(\cos p_y a+\cos p_z a) - \mu_{F_{1(2)}}
\label{dispersion_F}
\end{equation}
represent the normal-state electron spectra of the S and $\mathrm{F_{1(2)}}$ layers, respectively, and $a$ is the lattice constant. Each element of the matrix in Eq.~(\ref{eq:matrix_hamiltonian}) is a $4 \times 4$ matrix in the direct product of particle-hole and spin spaces, with the explicit layer-space structure shown.

The superconducting order parameter, taken to be real, is determined from the self-consistency equation
\begin{eqnarray}
\Delta =  \lambda  T \sum_{\omega_m}\int \frac{a^2 d^2 p}{(2\pi)^2} \frac{{\rm Tr}[\check {G}^{SS}(\bm p, \omega_m)\sigma_0\tau_-]}{2}, 
\label{eq:SC}    
\end{eqnarray}
where $\tau_- = (\tau_x - i \tau_y)/2$, and $\check {G}^{SS}$ denotes the $(2,2)$-element of $\check {G}(\bm p, \omega_m)$ in layer space—a $4 \times 4$ matrix in particle-hole and spin spaces representing the Green's function of the superconducting layer. In the calculations that follow, we take the superconducting order parameter of the isolated superconductor at $T = 0$ to be $\Delta_0 = 0.016 t_S$, which corresponds to a pairing constant of $\lambda = -1.33t_S$.

The electronic spectral density in the S layer for the parallel magnetization configuration ($\theta=0$) and spin $\sigma = \pm 1$ is given by
\begin{align}
A_{\sigma} (\varepsilon, \bm p) =-\frac{1}{\pi} 
{\rm Im}\left[\frac{{\rm Tr}[\check {G}^{R,SS}(\sigma_0+\sigma\sigma_z)(\tau_0+\tau_z)]}{4}\right],
\label{eq:DOS_P}    
\end{align}
where $\check {G}^{R,SS}$ is obtained from $\check {G}^{SS}$ via analytic continuation $i\omega_m \to \varepsilon+i\delta$, with $\delta$ a positive infinitesimal.

For the antiparallel magnetization configuration ($\theta=\pi$), where the magnetization vectors align along the $y$-axis, the spectral density becomes
\begin{align}
A_{\sigma} (\varepsilon, \bm p) =-\frac{1}{\pi} 
{\rm Im}\left[\frac{{\rm Tr}[\check {G}^{R,SS}(\sigma_0+\sigma\sigma_y)(\tau_0+\tau_z)]}{4}\right].
\label{eq:DOS_AP}    
\end{align}

We further examine the singlet and triplet correlations in the S layer, computed as 
\begin{eqnarray}
d_k (\omega_m) = \int \frac{d^2 p}{(2\pi)^2} \frac{{\rm Tr}[\check {G}^{SS} (\bm p, \omega_m) \sigma_k \tau_-]}{2},
\label{eq:Correlations}    
\end{eqnarray}
where $d_0$ represents the singlet correlation amplitude, and the vector $\bm d = (d_x, d_y, d_z)^T$ characterizes the triplet superconducting correlations. In the present system, $d_x$ becomes nonzero only for non-collinear magnetizations $\bm h_1$ and $\bm h_2$, signaling the non-collinear (triplet) spin-valve effect. The components $d_y$ and $d_z$, in contrast, are associated with the standard and inverse spin-valve effects, respectively.

\section{Gate-controlled standard and inverse spin-valve effects}
\label{sec:St_Inv}

\begin{figure}[tb]
	\begin{center}
		\includegraphics[width=70mm]{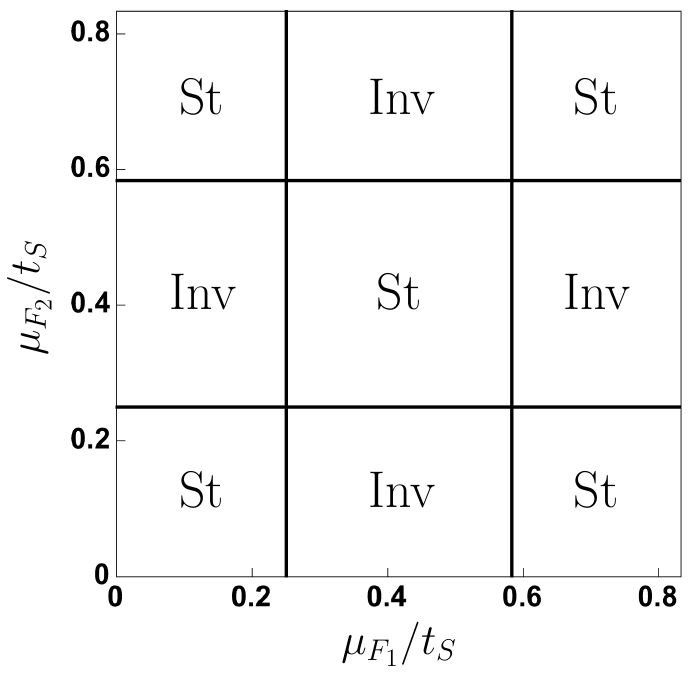}
		\caption{\textbf{Phase diagram for standard and inverse SVE.} The diagram delineates the regions in the $(\mu_{F_1}, \mu_{F_2})$ parameter space where the standard (St) and inverse (Inv) SVEs are realized.  Parameters used are $t_F = 1.25 t_S$, $t_{FS} = 0.017 t_S$, $\mu_S = 0.333 t_S$, and $h = 0.167 t_S$.}
		\label{fig:St_Inv}
	\end{center}
\end{figure}

\begin{figure}[tb]
	\begin{center}
		\includegraphics[width=85mm]{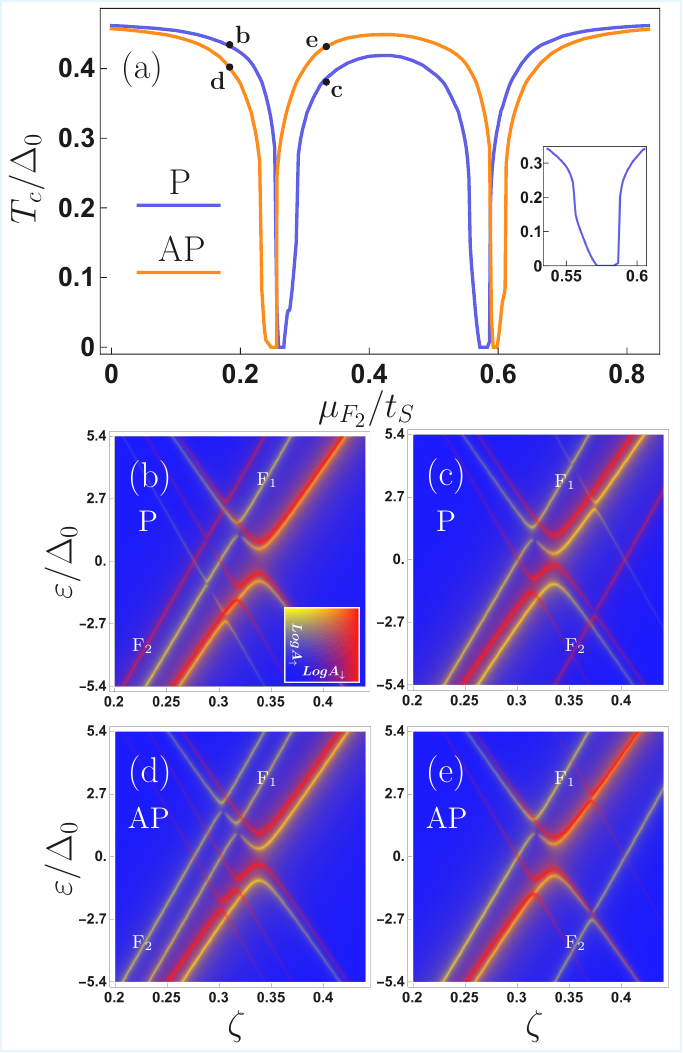}
		\caption{\textbf{$T_c$ and electronic spectra of the $\mathbf{F_1/S/F_2}$ spin valve.}
(a) $T_c$ vs.\ $\mu_{F_2}$ of the $\mathrm{F}_2$ layer for parallel (P, blue) and antiparallel (AP, orange) magnetization alignments.
Inset: zoom into the suppressed-superconductivity region for P.
(b)--(e) Spin-resolved spectral density.
$A_\uparrow(\zeta,\varepsilon)$ and $A_\downarrow(\zeta,\varepsilon)$ [$\zeta = -2(\cos p_y a + \cos p_z a)$] are overlaid.
Color code: blue, no states; yellow (red), pure spin-up (spin-down); orange, mixture [exact $\log A_{\uparrow,\downarrow}(\zeta,\varepsilon)$ values range from $-9$ to $3$, see inset to (b)].
Spectra are taken at points b--e marked in (a); branches of $\mathrm{F}_1$ and $\mathrm{F}_2$ layers are labeled.
Parameters: $T = 0.214\Delta_0$, $t_F = 1.25 t_S$, $t_{FS} = 0.017 t_S$, $\mu_S = 0.333 t_S$, $\mu_{F_1} = 0.55 t_S$, $h = 0.167 t_S$.
Specific values: (b),(d) $\mu_{F_2} = 0.183 t_S$; (c),(e) $\mu_{F_2} = 0.333 t_S$.}
		\label{fig:TcMu}
	\end{center}
\end{figure}

One of central findings of our work is the demonstration that switching between standard and inverse SVE can be achieved by electrically gating either ferromagnetic layer to modulate its on-site energy. Fig.~\ref{fig:St_Inv} presents a phase diagram mapping the regions in $(\mu_{F_1}, \mu_{F_2})$ parameter space where standard and inverse SVE are realized within our model framework. Fig.~\ref{fig:TcMu}(a) displays a characteristic dependence of $T_c$ on the on-site energy of one of the ferromagnetic layers for both parallel ($T_c^{P}$, blue) and antiparallel ($T_c^{AP}$, orange) magnetization configurations. The distinct behavior of the blue and orange curves reveals that the system exhibits standard SVE ($T_c^{AP}>T_c^{P}$) or inverse SVE ($T_c^{AP}<T_c^{P}$) depending on the value of $\mu_{F_2}$. Notably, near the boundaries separating standard and inverse SVE regions—where superconductivity suppression is most pronounced—we observe absolute SVE, characterized by either $T_c^{AP} = 0$ with $T_c^{P} \neq 0$ or vice versa. Taking $\Delta_0 \sim 10$ K $\sim 1$ meV, the full studied range corresponds to $\Delta \mu_{F_{1,2}} \sim 50$ meV, with absolute SVE regions spanning $\delta \mu_{F_{1,2}} \sim 1$ meV. This indicates that different SVE types, including absolute SVE, can be continuously tuned via gating.

The inset to Fig.~\ref{fig:TcMu}(a) shows an expanded view of the region of suppressed superconductivity for the parallel configuration. It reveals a nontrivial, cusp-like suppression of $T_c$, which signals a departure from conventional BCS-like behavior and indicates the presence of bistable and reentrant superconducting states as a function of temperature, as discussed in detail in Sec.~\ref{sec:OP_T}.

The gate-tunability of the SVE is fundamentally rooted in the electronic band hybridization that governs proximity effects in few-layer van der Waals heterostructures \cite{Bobkov2024_vdW}. To elucidate the underlying mechanism, Figs.~\ref{fig:TcMu}(b)-(e) present the spin-resolved electronic spectral density calculated for system parameters corresponding to points b-e in Fig.~\ref{fig:TcMu}(a). The spectral densities for parallel and antiparallel magnetization configurations were calculated using Eqs.~(\ref{eq:DOS_P}) and (\ref{eq:DOS_AP}), respectively. Within our simplified model framework, it is advantageous to represent the spectra as a function of $\zeta = -2(\cos p_y a + \cos p_z a)$ rather than along specific momentum directions. Hyperbolic spectral branches originate from the superconductor, while each ferromagnet contributes two spin-split linear branches, one of which is visible and labeled accordingly; the remaining ferromagnetic branches lie outside the displayed $(\zeta, \varepsilon)$ range.

Finite interlayer hopping $t_{FS}$ causes hybridization between superconducting and ferromagnetic branches, yielding a Zeeman splitting of the superconducting branches that is determined by the relative alignment of the superconducting and ferromagnetic spectral branches. Neglecting $t_{FS}$, the ferromagnetic branch follows $\xi_{F_{1,2}}^\sigma = t_F \zeta - \mu_{F_{1,2}} + \sigma h$ [see Eqs.~(\ref{eq:matrix_hamiltonian}) and (\ref{dispersion_F})], implying that the relative alignment between ferromagnetic ($F_{1,2}$) and superconducting branches is governed by the on-site energy $\mu_{F_{1,2}}$.

The spin splitting of superconducting branches can be quantified by the effective exchange field $h_{eff}$ induced in the S-layer by the ferromagnets. Qualitatively, $h_{eff}$ can be approximated as a superposition of contributions from individual ferromagnetic layers. Though rigorous calculations account for their mutual influence, this influence is generally weak across broad parameter ranges and negligible for the qualitative analysis of the standard and inverse SVE. In the limit of weak interlayer hopping $t_{FS} \ll |\xi_F^\sigma (\zeta_0)|$, with $\zeta_0$ determined by $\xi_S (\zeta_0)=0$, the effective exchange field induced by a ferromagnet in an S/F heterostructure takes the form \cite{Ianovskaia2024}:
\begin{align}
    h_{eff}=\frac{-ht_{FS}^2}{\xi_F^+(\zeta_0)\xi_F^-(\zeta_0)} = \frac{ht_{FS}^2}{h^2-[(\mu_S t_F/t_S)-\mu_F]^2}.
    \label{eq:heff}
\end{align}
Maximal spin splitting of superconducting branches occurs when $\xi_{F}^\sigma(\zeta_0) = 0$, i.e., when a superconducting branch crosses a ferromagnetic branch at the Fermi level $\varepsilon=0$ [though Eq.~(\ref{eq:heff}) becomes inaccurate in this regime]. Furthermore, the splitting changes sign as $\xi_F^\sigma(\zeta_0)$ changes sign, which occurs when a ferromagnetic branch passes through the superconducting branch upon variation of the corresponding on-site energy. This sign reversal underlies the SVE switching observed between points b-d and e-c in Fig.~\ref{fig:TcMu}(a). At point b, the $h_{eff}$ contributions from $F_1$ and $F_2$ partially cancel [Fig.~\ref{fig:TcMu}(b)], whereas at point c they add constructively despite {\it unchanged parallel magnetization alignment}. An analogous but reversed sequence applies to the antiparallel configuration [Figs.~\ref{fig:TcMu}(d)-(e)].

These findings remain robust against variations in model parameters, including the case of non-identical ferromagnets. Adjusting parameters such as $h_1$, $h_2$, $t_{F}$, $t_{S}$ or $\mu_S$ merely shifts the values of $\mu_{F_1}$ and $\mu_{F_2}$ at which the strongest hybridization occurs, thereby displacing the regions of maximal $T_c$ suppression in Fig.~\ref{fig:TcMu}(a) and shifting the boundaries between the standard and inverse SVE regimes in Fig.~\ref{fig:St_Inv}.  Moreover, more substantial modifications to the dispersion relation—such as the inclusion of longer-range hopping terms—likewise only shift the boundaries between the standard and inverse SVE regimes. This follows directly from Eq.~(\ref{eq:heff}), which shows that the effective field induced by each ferromagnet in the superconducting layer is determined solely by the electron energies at the crossing points of the electronic branches of the ferromagnet and superconductor spectra—that is, by the quantities $\xi_F^{\pm}(\zeta_0)$—and not by the specific form of the dispersion relation.

As already noted in Sec.\ref{sec:model}, the physics becomes more intricate when spin-orbit coupling is present in the superconductor. In this case, the orientation of each magnetization relative to the S/F interface also becomes a crucial parameter. We expect that our results regarding electrical controllability and switching between the standard and inverse spin-valve effects apply directly to the case where both magnetizations are oriented perpendicular to the interfaces (i.e., out of plane). Of course, the precise boundaries between the standard and inverse SVE regimes may shift because of the more complex band structures. For in-plane magnetizations, the same controllability and switching behavior is anticipated, though the magnitude of the spin-valve effect itself could be reduced owing to the smallness of the induced triplet correlations. A detailed investigation of the SVE amplitude in this case lies beyond the scope of the present work.

\section{Gate-controlled triplet spin-valve effect}

\label{sec:Triplet}

The triplet SVE—also referred to as the nonmonotonic SVE—manifests as a non-monotonic dependence of both the order parameter $\Delta$ and the critical temperature $T_c$ on the magnetization misorientation angle $\theta$. As noted in the Introduction, the characteristic minimum in $T_c(\theta)$ or $\Delta(\theta)$ arises from the generation of spin-triplet correlations with a $\bm d$ vector proportional to $\bm h_1 \times \bm h_2$. These correlations correspond to equal-spin triplet pairs when quantized along an axis parallel to the total effective exchange field in the superconductor, i.e., along \(\bm h_1+\bm h_2\) or \(\bm h_1-\bm h_2\). In conventional multilayer ferromagnets, such triplets are often termed long-range triplet correlations \cite{Bergeret2005}, as their amplitude decays much more slowly into the ferromagnet than that of opposite-spin pairs with $\bm d \parallel \bm h$. However, in the present monolayer system, the distinction between short- and long-range triplet components is no longer meaningful. The emergence of $\bm h_1 \times \bm h_2$ pairs opens an additional singlet-triplet conversion channel, further suppressing singlet superconductivity. These triplet correlations peak near $\theta = \pi/2$ resulting in the strongest suppression of superconductivity.

\begin{figure}[tb]
	\begin{center}
		\includegraphics[width=85mm]{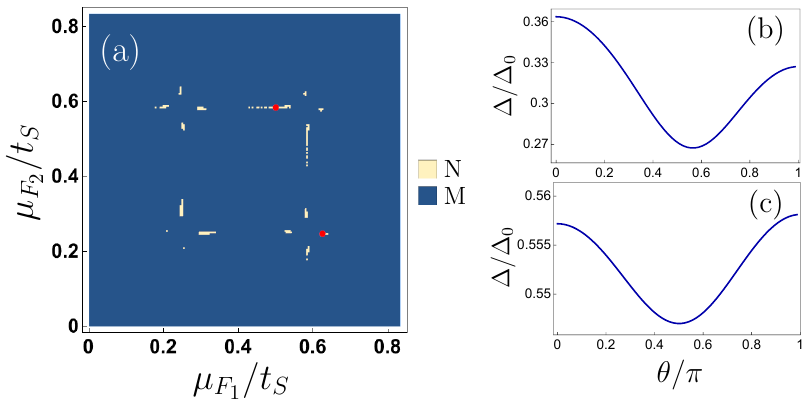}
\caption{\textbf{Non-monotonic angular dependence of the superconducting order parameter.} (a) Phase diagram indicating regions of monotonic (M) and non-monotonic (N) $\Delta(\theta)$ behavior. (b)-(c) Examples of non-monotonic $\Delta(\theta)$ dependencies corresponding to the red points in panel (a): (b) $\mu_{F_1} = 0.625 t_S$, $\mu_{F_2} = 0.247 t_S$; (c) $\mu_{F_1} = 0.5 t_S$, $\mu_{F_2} = 0.583 t_S$.  Other parameters are the same as in Fig.~\ref{fig:TcMu}.}
 \label{fig:Nonmon_OP}
	\end{center}
 \end{figure}

Figure~\ref{fig:Nonmon_OP}(a) identifies the regions in the $(\mu_{F_1},\mu_{F_2})$ plane where the triplet SVE is realized within our model. Representative examples of nonmonotonic $\Delta(\theta)$ are shown in Figs.~\ref{fig:Nonmon_OP}(b)–(c). Although the regions supporting the triplet SVE appear relatively narrow, their accessibility via continuous gate-voltage tuning of the on-site energies facilitates experimental observation.

\begin{figure}[tb]
	\begin{center}
		\includegraphics[width=85mm]{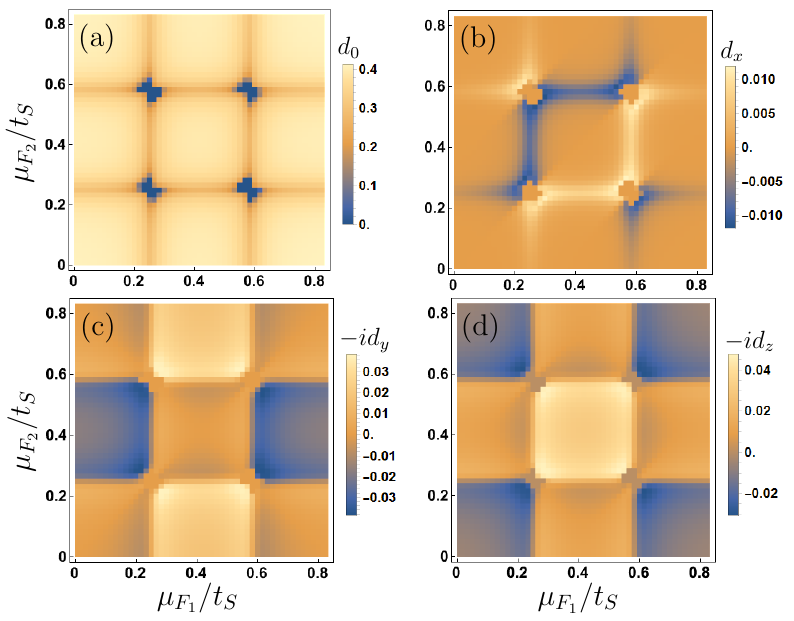}
\caption{{\bf Singlet and triplet superconducting correlations as functions of  $\bm {(\mu_{F_1}, \mu_{F_2})}$. } (a) Singlet correlations $d_0 (\omega_0)$ at the first Matsubara frequency, (b) triplet correlations $d_x(\omega_0)$, (c) triplet correlations $-i d_y (\omega_0)$, (d) triplet correlations $-i d_z (\omega_0)$. $\theta=\pi/2$. Other parameters are the same as in Fig.~\ref{fig:TcMu}.}
 \label{fig:Correlations}
	\end{center}
 \end{figure}

To gain further insight into the proximity effect in the vdW spin valve, we plot the phase diagrams of singlet ($d_0$) and triplet ($\bm d$) superconducting correlations in Figs.~\ref{fig:Correlations}(a)–(d). These results, calculated self-consistently via Eq.~(\ref{eq:Correlations}) at $\theta = \pi/2$, show that suppression of singlet correlations is most pronounced in regions where the electronic spectra of the S and one of the F layers hybridize strongly, i.e., where $\xi_{F}^\sigma(\zeta_0) = 0$. When this condition is met for both ferromagnetic layers simultaneously, singlet correlations and the order parameter are completely suppressed [blue regions in Fig.~\ref{fig:Correlations}(a)].
 
The triplet component proportional to $\bm h_1 \times \bm h_2$, which in our geometry aligns with the $x$-axis, is presented in Fig.~\ref{fig:Correlations}(b). This component reaches its maximum amplitude in regions of strong hybridization, consistent with suppression of singlet correlations. The $d_x$ correlations exhibit antisymmetry under the exchange $\mu_{F_1} \leftrightarrow \mu_{F_2}$, as required by symmetry considerations. Specifically, a $\pi$-rotation of the spin valve about the $z$ axis exchanges $\mu_{F_1}$ and $\mu_{F_2}$ and reverses the sign of $d_x$, leaving the exchange fields unchanged. The magnitude of $d_x$ is several times smaller than that of the other correlation components. In the limit of weak interlayer hopping $t_{FS} \ll (t_{S,F}, \mu_{S,F_{1,2}}, \Delta, T)$, the analytical expression for $d_x$ takes the form:
\begin{align}
&d_x  \approx 
\frac{4 \ t^4_{FS} \omega_m \Delta    (\bm h_1 \times \bm h_2)_x (\xi_{F_1}-\xi_{F_2})}{ (\Delta^2 + \xi^2_{S} + \omega_m^2)^2 } \times \nonumber \\
&\prod\limits_{\gamma=\pm1} \frac{\sqrt{h^2 + \xi_{F_1}\xi_{F_2} + \omega_m^2}}{\bigl[h^2-(\xi_{F_1}+ \gamma i\omega_m)^2\bigr]\bigl[h^2-(\xi_{F_2}+ \gamma i\omega_m)^2\bigr]} .
\label{eq:triplet_1}
\end{align}

Figs.~\ref{fig:Correlations}(c) and (d) display the triplet correlations $d_y$ and $d_z$, respectively. Away from the immediate vicinity of the strong hybridization regions ($\xi_{F}^\sigma(\zeta_0) = 0$), only one of the two components—$d_y$ or $d_z$—is nonzero. This behavior stems from the alignment of these triplet components with the total effective field $\bm h_s = \bm h_{eff,1} + \bm h_{eff,2}$ induced in the S layer. When the effective fields $\bm h_{eff,1(2)}$ induced by each of the $\rm F_{1,2}$ layers separately are both parallel or both antiparallel to $\bm h_{1(2)}$, the total field $\bm h_s$ lies along the $z$ axis. Conversely, if $\bm h_{eff,1}$ is antiparallel to $\bm h_1$ while $\bm h_{eff,2}$ is parallel to $\bm h_2$ (or vice versa), $\bm h_s$ aligns with the $y$ axis. It is worth noting that in F/S/F structures with conventional thin S films—which consist of many monolayers but have a thickness smaller than the superconducting coherence length—the net effective exchange field $\bm h_s = \bm h_{eff,1} + \bm h_{eff,2}$ is the only Zeeman-related effect of the F layers on the superconductor. In this case, the triplet proximity effect, which is governed by $\bm h_1 \times \bm h_2$, cannot arise. The triplet SVE emerges only for thicker films with $d \gtrsim \xi_{\rm cl}$, where the proximity effect with two ferromagnets can no longer be reduced to a simple vector sum of the exchange fields they induce.

In our case, approaching the regime of strongest superconductivity suppression [dark blue regions in Fig. ~\ref{fig:Correlations}(a)] leads to such a strong hybridization of the superconductor and ferromagnet spectra that the S layer can no longer be described as a superconductor in an effective field. Indeed, the electronic spectra of the system differ so markedly from those of the isolated superconductor and ferromagnets that the Zeeman splitting of the superconducting branches cannot be identified. Consequently, even for a superconducting monolayer, equal-spin pairs emerge, described by a component of the $\bm d$ vector proportional to $\bm h_1 \times \bm h_2$. This contribution arises from the cross–proximity effect of both ferromagnets simultaneously and gives rise to the triplet SVE.

As follows from Eq.~(\ref{eq:triplet_1}), modifying or complicating the dispersion relation—by adjusting the hopping parameters or including longer-range hopping—does not affect the qualitative conclusion regarding the existence and structure of the $d_x$ component responsible for the triplet SVE. This is because the expression involves only the electron energies in the normal state of the ferromagnet and superconductor.

The influence of SOC is less straightforward and requires careful consideration. Qualitatively, we offer the following assessment. If the magnetization vectors are constrained to rotate strictly within the plane of the interface, we do not expect any qualitative deviations from our predictions. However, if at $\theta = 0$ both magnetizations lie in the plane, but upon rotation one of them acquires an out-of-plane component, we anticipate an enhancement of the nonmonotonicity in $T_c(\theta)$. This enhancement arises because when a magnetization tilts out of the plane, the $h_{\rm{eff}}$ that suppresses superconductivity increases sharply. Conversely, if at $\theta = 0$ both magnetizations are perpendicular to the plane, then upon rotation, as one of them aligns with the plane, $h_{\rm{eff}}$ diminishes, thereby reducing its pair-breaking effect. This could lead to a weakening or even complete disappearance of the nonmonotonic behavior with minimum at $\theta = \pi/2$.

\section{Temperature dependence of the superconducting order parameter}

\label{sec:OP_T}

As discussed in the previous section, approaching the regime of strong superconductivity suppression means that the S layer can no longer be described as a superconductor in an effective field. In this regime, not only does the triplet SVE emerge, but a variety of exotic superconducting states also appear. Since this regime can be accessed via electrical control, the temperature dependence of the order parameter becomes highly tunable by gating, even for the parallel magnetization configuration ($\theta=0$). 

\begin{figure}[tb]
	\begin{center}
		\includegraphics[width=85mm]{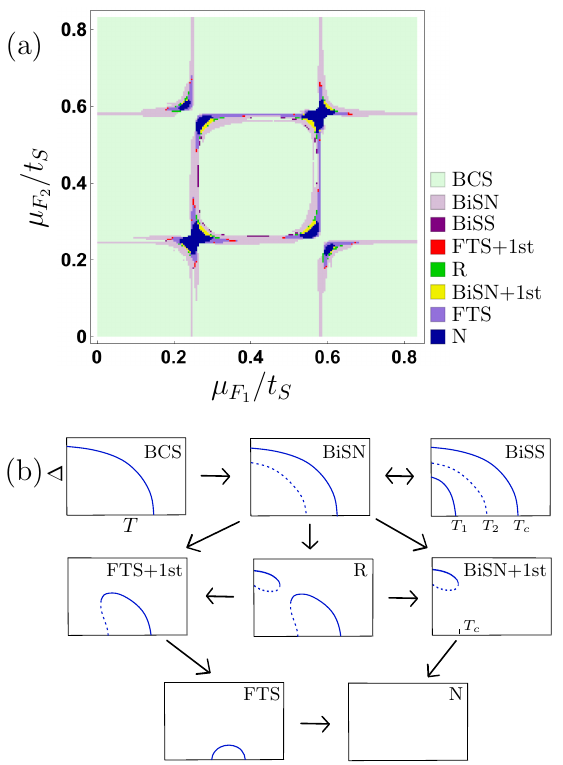}
\caption{(a) Phase diagram in the $(\mu_{F_1}, \mu_{F_2})$ plane identifying regions hosting distinct temperature-dependent behaviors of the order parameter. Parallel magnetic configuration. Color-coded areas correspond to the different regimes identified, see the legend. (b) Schematic illustration of the possible evolution of $\Delta(T)$, attainable by gating, from an unsuppressed BCS-like superconducting state to the fully suppressed normal state. Each curve qualitatively represents the characteristic temperature dependence within the corresponding phase identified in (a). All parameters except for $\mu_{F_{1,2}}$ are the same as in Fig.~\ref{fig:TcMu}.}
 \label{fig:Delta_T}
	\end{center}
 \end{figure}

Fig.~\ref{fig:Delta_T}(a) presents the complete phase diagram in the $(\mu_{F_1}, \mu_{F_2})$ plane, identifying regions where different types of temperature-dependent OP behavior emerge within our model. Depending on the values of $(\mu_{F_1}, \mu_{F_2})$, we distinguish: conventional BCS-like suppression (BCS); bistable states with two stable superconducting solutions (BiSS) or coexisting superconducting and normal states (BiSN); first-order transitions between superconducting and normal states (1st); reentrant superconductivity (R); and the emergence of superconductivity at finite temperature (FTS). 

Figure~\ref{fig:Delta_T}(b) illustrates all possible types of $\Delta(T)$ evolution accessible by varying the ferromagnetic on-site energies. Beginning from the conventional BCS regime and adjusting $(\mu_{F_1}, \mu_{F_2})$ via gating, superconductivity can be fully suppressed through various intermediate non-BCS states, as listed above. Schematic curves depict characteristic $\Delta(T)$ behaviors in each region; numerically calculated examples at specific $(\mu_{F_1}, \mu_{F_2})$ points are provided in Appendix~\ref{app:Delta_T}.

This entire variety of non-BCS states occurs in the vicinity of points where $\xi_{F_1}^\pm (\zeta_0) = \xi_{F_2}^\pm (\zeta_0) = 0$, corresponding to maximum spectral hybridization with both ferromagnets. In contrast, along lines where maximum hybridization occurs with only one ferromagnet [$\xi_{F_1}^\pm (\zeta_0) = 0$, $\xi_{F_2}^\pm (\zeta_0) \neq 0$], the BiSN phase predominantly appears, as expected for a superconductor in a Zeeman field \cite{Sarma1963}.

The phases listed above are classified according to the most important characteristic features. For example, for the BiSS phase the system has two stable superconducting solutions only at $T<T_1$, and at $T_1<T<T_2$ [see picture BiSS in Fig.~\ref{fig:Delta_T}(b)] two stable solutions of the system are superconducting and normal. In this sense  at $T_1<T<T_2$ the BiSS-phase is identical to the BiSN-phase. Some regions of the  $(\mu_{F_1}, \mu_{F_2})$ plane exhibit a combination of two phases according to our classification. For example, in the picture BiSN+1st we can see the BiSN phase at $T<T_c$, but the transition from the superconducting state to the normal state at $T=T_c$ is a first-order transition in contrast to the picture BiSN, where the transition is of second order. Additionally, in Appendix~\ref{app:Delta_T}, the region of the phase diagram around one of the points of maximum order parameter suppression—where unconventional superconducting states emerge—is presented at a higher resolution.

\begin{figure}[tb]
	\begin{center}
		\includegraphics[width=85mm]{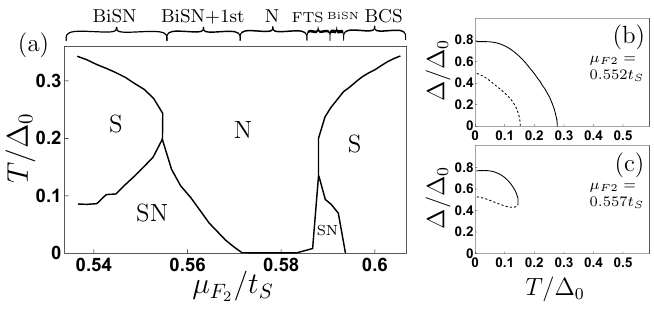}
\caption{(a) Phase diagram in the $(\mu_{F_2},T)$ plane for the region of suppressed superconductivity, corresponding to the area magnified in the inset of Fig.~\ref{fig:TcMu}(a). The top of the panel indicates the specific phases realized within each interval of $\mu_{F_2}$. Lines separate regions where only the superconducting state (S), both superconducting and normal states (SN), or only the normal state (N) are stable. (b)–(c) Representative examples of the temperature evolution of the order parameter $\Delta(T)$, as discussed in the main text.}
 \label{fig:Phase_Diag}
	\end{center}
 \end{figure}

Given the identification of the possible superconducting phases realizable via gating in our spin-valve system, we can now interpret the nontrivial, cusp-like suppression of $T_c$ presented in the inset of Fig.~\ref{fig:TcMu}(a). The phase diagram in the $(\mu_{F_2},T)$ plane for the region of suppressed superconductivity is shown in Fig.~\ref{fig:Phase_Diag}(a). At small $\mu_{F_2}$, the system resides in a bistable superconducting/normal (BiSN) state. This implies that at low temperatures, both superconducting  and normal states are thermodynamically stable [SN region in Fig.~\ref{fig:Phase_Diag}(a)], while at higher temperatures only the superconducting phase is stable [S region in Fig.~\ref{fig:Phase_Diag}(a)]. An example of $\Delta(T)$ curve for this regime is displayed in Fig.~\ref{fig:Phase_Diag}(b). Owing to the non-BCS temperature dependence of the order parameter, the superconducting-to-normal transition does not occur directly but proceeds through an intermediate phase combining bistability and a first-order transition (BiSN+1st), as shown in Fig.~\ref{fig:Phase_Diag}(b). The cusp in the $T_c(\mu_{F_2})$ dependence highlighted in the inset of Fig.~\ref{fig:TcMu}(a) is a direct consequence of this transition between different superconducting phases.

Various forms of non-BCS superconductivity have been previously proposed. Notable examples include bistable BiSN states and first-order phase transitions in thin-film S/F heterostructures \cite{Bobkova2014}, thin superconducting films under parallel magnetic fields \cite{Sarma1963}, reentrant superconductivity in S/F bilayers \cite{Fominov2002,Zdravkov2006,Zdravkov2010,Sidorenko2017} and F/S/F trilayers \cite{Lenk2017,Zdravkov2013_Reentrant,Antropov2013,Kehrle2012}, and the emergence of superconductivity at finite temperatures under non-equilibrium conditions \cite{Bobkova2017_thermospin}. The fundamental significance and experimental advantage of our F/S/F trilayer lie in its capacity to host all these exotic superconducting states within a single, fully controllable system. Crucially, these states can be accessed through smooth adjustment of external gate voltages without imposing stringent constraints on the system parameters.

\section{Conclusions}

In summary, we have demonstrated that van der Waals F$_1$/S/F$_2$ trilayers provide a highly versatile platform for implementing and controlling various types of superconducting spin-valve effects through electrostatic gating. Our theoretical investigation, based on a minimal tight-binding model and Green's function approach, reveals several key findings:

First, the standard, inverse, and triplet spin-valve effects can all be realized within the same heterostructure by tuning the onsite energies of the ferromagnetic layers via gate voltages. The transitions between these regimes are governed by the hybridization between superconducting and ferromagnetic electronic spectra, which determines the effective exchange field induced in the superconducting layer.

Second, gate control extends beyond the spin-valve functionality to the fundamental temperature dependence of the superconducting order parameter. Even for parallel magnetization alignment, we observe a rich variety of non-BCS behaviors including reentrant superconductivity, bistable states, first-order phase transitions, and the emergence of superconductivity at finite temperatures.

The accessibility of different spin-valve regimes and the exotic superconducting states through smooth gate voltage adjustments, without requiring stringent parameter constraints, highlights the unique advantages of van der Waals heterostructures for superconducting spintronics. The predicted effects are experimentally relevant, given the established capabilities for electrostatic gating in few-layer materials and the recent progress in fabricating high-quality van der Waals heterostructures.

Our work establishes gate-tunable van der Waals spin valves as promising building blocks for future superconducting spintronic devices, offering unprecedented electrical control over superconducting phenomena and opening avenues for exploring the rich physics of proximity effects in low-dimensional systems.

\label{sec:conclusions}

\begin{acknowledgments}
A.S.I., G.A.B. and I.V.B. acknowledge the support from Theoretical Physics and Mathematics Advancement Foundation “BASIS” via the Project  No. 23-1-1-51-1. The calculations of the gate-controllable spin valve effect were supported by the Russian Science Foundation via the Project No. 24-12-00152. The study of the gate-controlled temperature dependence of the order parameter has been performed under the support by Grant from the ministry of science and higher education of the Russian Federation No. 075-15-2025-010.  
\end{acknowledgments}

\appendix

\section{Derivation of the Gor'kov equation for a F/S/F vdW heterostructure}

\label{sec:App_Gorkov}

In this appendix we present key steps of the derivation of the Gor'kov equation for the trilayer heterostructure. The Green’s function Eq.~(\ref{eq:Green_Gorkov}) obeys the following equation:
\begin{eqnarray}
\frac{d \check{G}_{\bm i \bm j}}{d \tau_1} = - \delta(\tau_1 -\tau_2) \delta_{\bm i \bm j} - \tau_z \langle T_{\tau } \frac{d \check{\psi_{\bm i}} (\tau_1)}{d \tau_1} \check{\psi^{\dag}_{\bm j}} (\tau_2) \rangle.
\label{eq:Green_derivative}
\end{eqnarray}
For the system described by Hamiltonian (\ref{ham}) the Heisenberg equation of motion for spinor $\check \psi_{\bm i}$ takes the form:

\begin{figure*}[!tbh]
	\begin{center}
		\includegraphics[width=\textwidth]{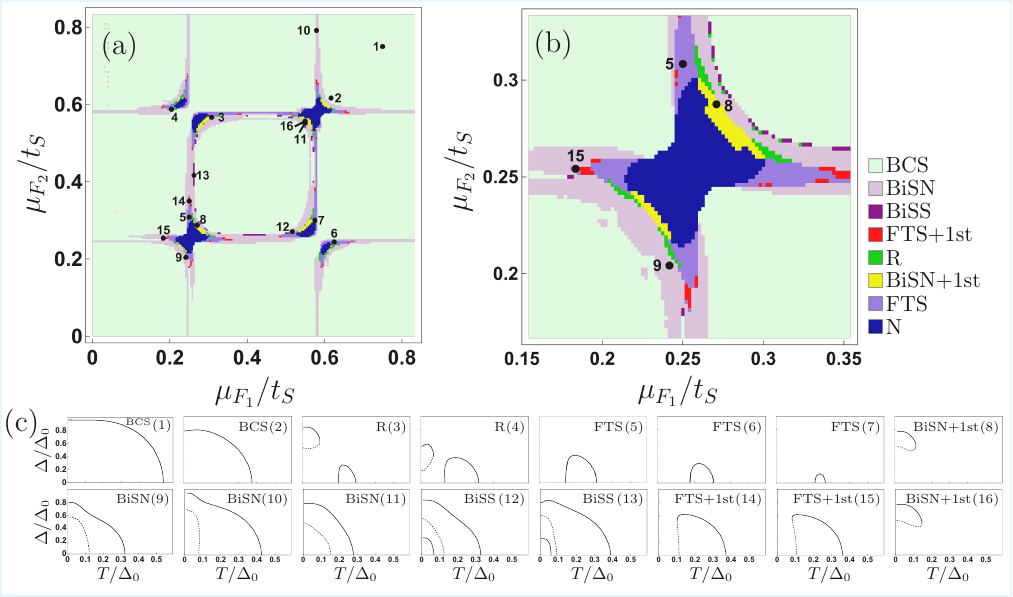}
\caption{(a)  Phase diagram in the $(\mu_{F_1}, \mu_{F_2})$ plane identifying regions hosting distinct temperature-dependent behaviors of the order parameter. (b) The region of the phase diagram around one of the points of maximum order parameter suppression presented in a higher resolution. (c) Specific examples of $\Delta(T)$ behavior corresponding to $(\mu_{F_1}, \mu_{F_2})$-points marked in panel (a) by numbers (1) $\mu_{F_1}=0.75 t_S, \mu_{F_2}=0.75 t_S$, (2) $\mu_{F_1}=0.617 t_S, \mu_{F_2}=0.617 t_S$, (3) $\mu_{F_1}=0.308 t_S, \mu_{F_2}=0.567 t_S$, (4) $\mu_{F_1}=0.204 t_S, \mu_{F_2}=0.588 t_S$, (5) $\mu_{F_1}=0.25 t_S, \mu_{F_2}=0.308 t_S$, (6) $\mu_{F_1}=0.625 t_S, \mu_{F_2}=0.244 t_S$, (7) $\mu_{F_1}=0.575 t_S, \mu_{F_2}=0.3 t_S$, (8) $\mu_{F_1}=0.271 t_S, \mu_{F_2}=0.288 t_S$, (9) $\mu_{F_1}=0.242 t_S, \mu_{F_2}=0.204 t_S$, (10) $\mu_{F_1}=0.579 t_S, \mu_{F_2}=0.792 t_S$, (11) $\mu_{F_1}=0.55 t_S, \mu_{F_2}=0.552 t_S$, (12) $\mu_{F_1}=0.517 t_S, \mu_{F_2}=0.271 t_S$, (13)  $\mu_{F_1}=0.263 t_S, \mu_{F_2}=0.417 t_S$, (14)  $\mu_{F_1}=0.25 t_S, \mu_{F_2}=0.35 t_S$, (15) $\mu_{F_1}=0.183 t_S, \mu_{F_2}=0.254 t_S$,  (16) $\mu_{F_1}=0.55 t_S, \mu_{F_2}=0.557 t_S$. Other parameters are the same as in Fig.~\ref{fig:TcMu}.}
 \label{fig:DT_Examples}
	\end{center}
 \end{figure*}

\begin{widetext}
\begin{align}
\frac{d \check{\psi_{\bm i}}}{d \tau} = [\hat{H}, \check{\psi_{\bm i}}]  = \hat M \check{\psi_{\bm i}} = 
\left(
\begin{matrix} t_F \hat{t}\tau_z + \mu_{F_1}\tau_z - \bm h_1 \bm \sigma  & t_{FS} \tau_z & 0   \\ t_{FS} \tau_z & t_S \hat{t} \tau_z + \mu_S \tau_z -  \check{\Delta}   &  t_{FS} \tau_z \\  0 & t_{FS}\tau_z  & t_F \hat{t} \tau_z + \mu_{F_2} \tau_z - \bm h_2 \bm \sigma 
\end{matrix}
\right) \check{\psi_{\bm i}} \label{eq:Heisenberg_equation} 
\end{align}
\end{widetext}
where $\check{\Delta}=\Delta\tau_{+} + \Delta^{*} \tau_{-}$ with $\tau_{\pm} = (\tau_x\pm i \tau_y)/2$. The operator $\hat{t}$ acts on the spinor $\check{\psi_{\bm i}} $ in the following way:
\begin{equation}
\hat{t} \check{\psi_{\bm i}} = \sum\limits_{\langle \bm i \bm j \rangle} \check{\psi_{\bm j}} = \sum\limits_{\langle \bm a \rangle} \check{\psi}_{\bm i + \bm a}.
\end{equation}
Here $\bm a \in \{0, \pm \bm a_y, \pm \bm a_z \}$ are the basis vectors in the plane of the layers. Substituting Eq.~(\ref{eq:Heisenberg_equation}) into Eq.~(\ref{eq:Green_derivative})  we obtain:
\begin{align}
G_{\bm i}^{-1} \check{G}_{\bm i \bm j} & (\tau_1-\tau_2) = \delta_{\bm i \bm j} \delta(\tau_1-\tau_2), \\
& G_{\bm i}^{-1} = (\hat{M}  - \frac{d }{d \tau_1})\tau_z.
\end{align}
The operator $\hat{t}$ acts on the Green's function $\check{G}_{\bm i \bm j}$ in the following way:
\begin{equation}
\hat{t} \check{G}_{\bm i \bm j} = \sum\limits_{\bm a} \check{G}_{\bm i + \bm a, \bm j}
\end{equation}
Further we introduce the Fourier transform of the Green’s function as defined by Eq.~(\ref{eq:mixed}). The term $\hat{j} \check{G}_{\bm i \bm j}$ in the momentum representation takes the form:
\begin{eqnarray}
\sum\limits_{\bm a} \int d^2 \bm r e^{-i \bm p (\bm i - \bm j)} \check{G}_{\bm i + \bm a, \bm j} = \nonumber \\
2 \check{G}(\bm p, \tau) [\cos(a_y p_y) + \cos(a_z p_z)],
\end{eqnarray}
where $\tau = \tau_1-\tau_2$. Then, expanding the Green’s function $\check{G}(\bm p, \tau)$ over fermionic Matsubara frequencies, one can obtain the Gor’kov equation for Green’s function:
\begin{align}
& G_{\bm p}^{-1} (\omega_m) \check{G} (\bm p, \omega_m) = 1, \\
& G_{\bm p}^{-1} (\omega_m) =  (\hat{M}_{\bm p}  + i \omega_m )\tau_z,
\end{align}
\begin{eqnarray}
\hat{M}_{\bm p} =  \left(\begin{matrix} M^{F_1}_{\bm p} & t_{FS}  &0 \\ t_{FS}  &M^{S}_{\bm p} &  t_{FS} \\  0 &t_{FS}  & M^{F_2}_{\bm p}  \\\end{matrix}\right)\tau_z,
\end{eqnarray}
\vspace{-0.4cm}
\begin{eqnarray}
    \begin{aligned}
M^{F_{1(2)}}_{\bm p} = 2 t_{F} [\cos(a_y p_y) + \cos(a_z p_z)] + \\
+ \mu_{F_{1(2)}} 
 - \bm h_{1(2)} \bm \sigma \tau_z,
     \end{aligned}
\end{eqnarray}
\vspace{-0.4cm}
\begin{eqnarray}
    \begin{aligned}
 M^{S}_{\bm p} = 2 t_S [\cos(a_y p_y) + \cos(a_z p_z)] +\\
 + \mu_S - \check{\Delta} \tau_z .
    \end{aligned}
\end{eqnarray}

\section{Temperature dependence of the superconducting OP: numerical results}
\label{app:Delta_T}

In Sec.~\ref{sec:OP_T}, schematic illustrations were provided to demonstrate the characteristic temperature behavior of the order parameter in different phases. In Fig.~\ref{fig:DT_Examples}, we present results of precise numerical calculations that confirm the expected behavior of the order parameter across these phases. Specifically, Fig.~\ref{fig:DT_Examples}(a) shows a phase diagram identical to that in Fig.~\ref{fig:Delta_T}(a), while Fig.~\ref{fig:DT_Examples}(b) presents a higher-resolution view of one of the regions where the superconductor hybridizes most strongly with both ferromagnets. In Fig.~\ref{fig:DT_Examples}(c), we display numerically calculated examples of $\Delta(T)$ at selected $(\mu_{F_1}, \mu_{F_2})$ points. The calculated curves for the corresponding phases exhibit qualitatively the same behavior as the schematic curves shown in Fig.~\ref{fig:Delta_T}(b). In addition, more subtle features of $\Delta(T)$ behavior are observed at specific points in the $(\mu_{F_1}, \mu_{F_2})$ plane. For instance, at the lowest temperatures the BCS-like behavior can be preceded by a small region where $\Delta(T)$ grows with increasing temperature—this occurs, for example, at points 2, 4, and 13.

\bibliography{vdW_FSF}

\end{document}